\documentstyle[12pt]{article}
\begin{document}
\pagestyle{empty}
\begin{flushright}
{CERN-TH/2000-217}
\end{flushright}
\vspace*{5mm}
\begin{center}
{\bf $O(d,d)$-INVARIANT COLLAPSE/INFLATION FROM COLLIDING SUPERSTRING WAVES} \\
\vspace*{1cm} 
{\bf V. Bozza$^{ab*}$  and  G.
Veneziano$^{c**}$ } \\
\vspace{0.3cm}
$^a$ Dipartimento  di Scienze Fisiche ``E.R. Caianiello'', 
Universit\`a di Salerno, Italy \\
$^b$ Istituto Nazionale di Fisica Nucleare, Sezione di   
Napoli, Naples, Italy \\
$^c$ Theoretical Physics Division, CERN, Geneva,  
Switzerland. \\
\vspace*{2cm}  
{\bf ABSTRACT} \\ \end{center}
\vspace*{5mm}
\noindent
Generalizing previous work, we study the collision of massless 
superstring plane waves in $D$ space-time   dimensions within an
explicitly $O(D-2,D-2)$-invariant set of field equations. We discuss
some general properties of the solutions, showing in particular that   they 
always lead to the formation of a singularity in the future.  Using the 
above symmetry, we obtain entire classes  of new analytic solutions with 
non-trivial metric, dilaton and   antisymmetric field, and discuss some of 
their properties of specific relevance to string cosmology.
\vspace*{5cm} 
\noindent 
\rule[.1in]{16.5cm}{.002in}

\noindent
$^{*)}$ valboz@sa.infn.it \\
$^{**)}$ Gabriele.Veneziano@cern.ch

\begin{flushleft} CERN-TH/2000-217 \\
July 2000
\end{flushleft}
\vfill\eject

\setcounter{page}{1}
\pagestyle{plain}

\section{Introduction}

One of the most debated issues in pre-big bang (PBB) cosmology \cite{PBB}  
concerns its initial conditions \cite{TW,BDV}. How fine--tuned should these 
be in order to generate a Universe resembling ours?  The principle of
asymptotic past   triviality (APT) \cite{BDV} amounts to saying that, far 
enough into the past, the Universe was well described by a {\it generic}, 
perturbative solution of the low-energy,  tree-level string effective
action. In other words, the very early Universe is   assumed to lie deeply 
inside the low-curvature, small--coupling regime. Under this assumption, 
for a critical superstring theory admitting the trivial Minkowski vacuum 
order by order in perturbation theory, the early Universe can be
represented as a random superposition (a chaotic sea) of massless waves,  
propagating  in all directions and with all (sub-string scale)
frequencies. The particle content of the waves should represent {\it all} 
the massless degrees of freedom of superstring theory, i.e. {\it each one} 
of the possible marginal deformations of the two-dimensional CFT in
trivial space-time. This can be hardly called a fine-tuned initial state!

Understanding the evolution of such a Universe is clearly {\it not} a
simple task. However, we know that, as long as the tree-level low-energy 
approximation remains valid, the classical field dynamics is both scale- 
and dilaton--shift--invariant since the string and Planck scales simply 
sit as overall factors in front  of the action. General arguments suggest 
that, while most of the time these waves will just  propagate linearly and 
independently, occasionally, through positive interference, overdense
regions will form and, thanks to gravitational instability, will  lead to 
gravitational collapse.  It was  argued \cite{BDV} that the interiors of  
sufficiently large and weakly coupled collapsing regions could give birth 
to Universes that could resemble our own. The above-mentioned classical 
symmetries guarantee that the scale of collapse will itself be a
stochastic variable, whose distribution will be  related to that of the 
original   distribution of wave lengths. Since such a distribution
naturally  contains arbitrarily long (though not arbitrarily short)
wavelengths, it looks very likely that large enough black holes would form 
with non-vanishing probability.

Notwithstanding the appeal of these general arguments, it would be very 
instructive   to see them at work in some explicitly soluble model. In 
\cite{BDV} this was partly done, in $D=4$ for the spherically symmetric 
case, through use of some powerful results obtained by Christodoulou
\cite{Chr} over more than a decade. More recently,   exact analytic
solutions were constructed by Feinstein, Kunze and V{\'a}zquez--Mozo  
(hereafter
FKV) \cite{FKV}, who replaced the chaotic sea of waves by two colliding,  
homogeneous, planar-fronted waves. In \cite{FKV} the  waves have infinite 
fronts and thus always lead to collapse  on space-time scales $L$ that are 
inversely related to the energy density  (i.e. energy per unit of
transverse area) in the waves:
\begin{equation}
L \sim (G_N \rho_1 G_N
\rho_2)^{-1/2}\;.
\label{focald}
\end{equation}
While FKV  only deal with $D=4$, we will
show that  the above  result  holds for any $D$. In the following we shall 
denote by $d = D-2$ the number of transverse coordinates.

Although the case of infinite fronts is obviously an idealization,
causality arguments lead to the conclusion that  collapse takes   place 
even for finite-front waves, provided the typical transverse extension $R$ 
of the waves  is larger than $L$, i.e. if:
\begin{equation}
R > L \sim \left( G_N
\rho_1 G_N \rho_2\right)^{-1/2} = R^d \left( G_N E_1 G_N E_2\right)^{-1/2} = 
\frac{R^{d}}{ G_N \sqrt{s}} = R_s ( \frac{R}{R_s} )
^{d},
\end{equation}
i.e. provided $R_s
\equiv (G_N \sqrt s)^{\frac{1}{d-1}} > R$, as naively expected.

 Other limitations of the FKV paper are that it deals only with
gravitational and dilatonic waves and that, as we mentioned, it is
restricted to $D=4$. It has been pointed out recently \cite {DH1} that 
the approach to the cosmological singularity  (which appears as the $r=0$ 
singularity inside the collapsing region) depends strongly both on the 
dimensionality of space and on the presence of  other fields, in
particular of the various $p$-forms that string theory possesses. It was 
argued, in particular, that turning on all the forms present in any
consistent string (or M) theory, changes the monotonic Kasner-like
behaviour of the gravi-dilaton system into an ever-oscillating behaviour \`a 
la BKL \cite{BKL}.

As a first step in the direction of overcoming the two limitations of the 
work by FKV, we shall extend it both to an arbitrary number of
(non-compact) dimensions and to the presence of the Kalb--Ramond
$B_{\mu\nu}$ field. Keeping the exact planar symmetry allows (see section 
2)  for an explicitly $O(d,d)$-invariant formulation of the dynamics in 
the string frame. While several properties of the solutions can be
discussed in the general case (section 3), we have only been able, so far, 
to construct explicit analytic solutions  in the case of $B_{\mu\nu}=0$ 
and arbitrary dimensions (sections 4, 5), or when the   $B_{\mu\nu}$ field is 
generated through $O(d,d)$-transformations (section 6). This will suffice, 
however,  to address some of the issues raised in   Ref. \cite{DH1}.

\section{The $O(d,d)$-invariant field equations}

As explained in the introduction, we start with  the low-energy,
tree-level effective action of critical superstring theory in the string 
frame. Up to a classically irrelevant overall factor
\begin{equation}
S=\int {dx}^D\sqrt{-g} e^{-\phi}\left({\rm R}+ \partial_ \mu \phi
\partial^\mu \phi -\frac{1}{12}H_{\mu\nu\rho}
H^{\mu\nu\rho}\right), \label{String action}
\end{equation}
where
$\phi$ is the dilaton, ${\rm R}$ is the curvature scalar for the metric 
tensor $g_{\mu\nu}$, and $H_{\alpha\beta\gamma}$ is the field strength 
for the antisymmetric field
\begin{equation}
H_{\mu\nu\rho}=\partial_\mu B_{\nu\rho}+ \partial_\rho
B_{\mu\nu}+ \partial_\nu B_{\rho\mu}.
\end{equation}

The problem we wish to study is the head-on collision of two
infinite plane--front waves, which, without loss of generality, will be 
taken to move   along the $x^1$ axis. Assuming symmetry under translations 
along the $d = D-2$ spatial directions orthogonal to $x^1$  ($x^i$ with $i 
= 2,\dots D-1$), it is obvious  that the problem is endowed with $d$
abelian isometries. In a convenient coordinate frame, there will be no 
dependence upon the ``transverse"   coordinates $x^i$, and thus, according 
to general arguments \cite{MV1,Sen},  we expect to have an exact $O(d,d)$ 
invariance of the classical field   equations. The symmetry of the field 
equations (not to be confused with a true quantum symmetry) is most easily 
exhibited in terms of the invariance of a ``reduced action" living in the 
non-trivial (here two-dimensional) subspace. In the case of
$O(D-1,D-1)$-symmetry (homogeneous Bianchi-I cosmologies) this was done in 
\cite{MV1,GV} and led to a continuous  extension of scale-factor duality 
of Bianchi-I cosmologies \cite{Tseytlin}.

For the present purposes, we will  adapt to our case a general result  by 
Maharana and Schwarz \cite{Mah Sch}, and write the reduced action   coming 
from (\ref{String action}) as:
\begin{equation}
S=\int
{dx^0}{dx^1}\sqrt{-g} e^{-\overline{\phi}} \left[ R+ \partial_ \alpha
\overline{\phi} \partial^\alpha \overline{\phi}
+\frac{1}{8}\mathrm{Tr}\left(\partial_\alpha M^{-1} \partial^\alpha M  
\right) \right],
\label{O(d,d) action}
\end{equation}
whose notation we shall now explain.

  The equations of motion allow the  metric $g_{\mu\nu}$ to be taken
block-diagonal, with blocks given by $g_{ij}$ and $g_{\alpha\beta}$  
where roman
indices ($i,j,\dots$) will span the components of the tensors from 2 to 
$d+1$ while the indices $\alpha$ and $\beta$ take the values 0 and 1. The 
explicit metric and curvatures appearing in (\ref{O(d,d) action}) only 
refer to the latter.

We arrange the components $g_{ij}$ in a $d$-dimensional matrix $G$. The 
matrix $B$ will contain the components $B_{ij}$ of the antisymmetric field 
while the remaining components of $B$ are set to $0$. $M$ is then the
$2d$-dimensional matrix defined by:
\begin{equation} M=\left(
\begin{array}{cc}
G^{-1} & -G^{-1}B \\
BG^{-1} & G-BG^{-1}B \end{array}
\right).
\end{equation}
Finally, the shifted dilaton is defined by:
\begin{equation}
\overline{\phi}=\phi-\frac{1}{2}\log \det G + {\rm
constant}\;,  \label{phibar def}
\end{equation}
where the constant will be
conveniently fixed later.

The reduced action (\ref{O(d,d) action})  is manifestly invariant under 
the transformations
\begin{eqnarray}
&& g_{\alpha\beta}\rightarrow g_{\alpha\beta}, \\
&& \overline{\phi}\rightarrow \overline{\phi}, \\
&& M \rightarrow
\Omega^T M \Omega ,
\end{eqnarray}
where $\Omega$ denotes a global
$O(d,d)$ transformation
\begin{equation}
\Omega^T \eta \Omega=\eta,
\label{O(d,d) definition}
\end{equation}
with $\eta$ the $O(d,d)$ metric
in off-diagonal form
\begin{equation}
\eta=\left( \begin{array}{cc}
0 & I_d \\
I_d & 0 \end{array}
\right).
\end{equation}
$M$ itself belongs to
$O(d,d)$ and, being symmetric,  satisfies
\begin{equation}
M\eta M=\eta \; , \;\; {\rm i. e.} \; \;  M^{-1} = \eta M \eta.
\end{equation}
The above
equations make the check of   $O(d,d)$-invariance trivial.

Let us now derive the (manifestly $O(d,d)$-invariant) field equations from 
 (\ref{O(d,d) action}).   Varying the action with respect to the shifted 
dilaton gives
\begin{equation}
R+2g^{\alpha \beta}D_\alpha D_\beta \overline{\phi}
-g^{\alpha \beta} \partial_\alpha \overline{\phi} \partial_\beta
\overline{\phi}
+\frac{1}{8}g^{\alpha\beta}\mathrm{Tr}\left(\partial_\alpha M^{-1}
\partial_\beta
M\right)=0 \label{phi equation},
\end{equation}
while varying with respect
to the $2$-metric  $g_{\alpha\beta}$ provides
\begin{equation}
R_{\alpha \beta} +
D_\alpha D_\beta \overline{\phi}+  
\frac{1}{8}\mathrm{Tr}\left(\partial_\alpha M^{-1}
\partial_\beta M\right)=0. \label{Einstein}
\end{equation}

Combining the trace of the latter equation with the previous one gives a 
simple equation for $\overline{\phi}$ which will play an important   role 
later:
\begin{equation}
D^\alpha D_\alpha {\rm exp}(- \overline{\phi}) = 0 .
\label{expphibar}
\end{equation}

The variation with respect to $M$ must be performed carefully because of 
the constraints coming from its definition. Following \cite{MV1}, we  take 
them into account by writing
\begin{equation}
\delta M=\Omega^T M \Omega-M ,
\label{Transformation for M}
\end{equation}
with $\Omega=1+\epsilon$ in
$O(d,d)$. Expanding to first order in $\epsilon$,  we obtain
\begin{equation}
\partial_\alpha \left(
e^{-\overline{\phi}}\sqrt{-g}g^{\alpha\beta} M^{-1} \partial_\beta M  
\right)=0.
\label{M equation}
\end{equation}

Equations. (\ref{phi equation}), (\ref{Einstein}), (\ref{expphibar}),
(\ref{M equation})
compose the full set we wish to solve. Before doing so, let us
write them more explicitly by going to the conformal gauge for
the 2-metric
\begin{equation}
g_{\alpha\beta} = e^F \eta_{\alpha\beta}
\end{equation}
and by working with ``light-cone" coordinates:
\begin{equation}
u = \frac{x^0-x^1}{\sqrt{2}}, \; \; \;v =
\frac{x^0+x^1}{\sqrt{2}}.
\end{equation}
Equations (\ref{phi
equation}), (\ref{Einstein}), (\ref{expphibar}), (\ref{M equation})  
then simply
become:
\begin{eqnarray}
&& \partial_u\partial_v {\rm exp}(-
\overline{\phi}) = 0, \,\, {\rm   i.e.} \,\,  \partial_u\partial_v
\overline{\phi} = \partial_u \overline{\phi} \partial_v \overline{\phi} \; 
; \label{phibar} \\
&& \partial_u \left( e^{-\overline{\phi}} M^{-1}
\partial_v M \right) + \partial_v \left( e^{-\overline{\phi}} M^{-1}
\partial_u M \right) = 0 \; ; \label{Mevol}\\
&&   \partial_u^2
\overline{\phi} - \partial_u  F  \partial_u    \overline{\phi} +
\frac{1}{8}{Tr}\left(\partial_u M^{-1} \partial_u M \right) = 0 , {\rm 
same} \; {\rm with} \; u   \rightarrow v \label{Vir};\\
&&   \partial_u
\partial_v \overline{\phi} -  \partial_u \partial_v F +
\frac{1}{8}{Tr}\left(\partial_u M^{-1} \partial_v M \right)=0.
\label{Fevol}
\end{eqnarray}

Note that the two equations (\ref{Vir}) have the form of Virasoro
constraints of a two--dimensional CFT, while the other equations are of the
evolution
type. Furthermore, a straightforward, though not completely trivial,   
calculation shows that the integrability condition for eqs. (\ref{Vir}) 
holds and that eq. (\ref{Fevol}) is just a consequence of the previous 
ones. Our discussion will thus be based on solving the set of equations 
(\ref{phibar})--(\ref{Vir}).

\section {General properties of the solutions}

 In this section we will derive some general features of the solutions, which
will be useful for the  more detailed investigations to be described  
later on.

The two colliding waves are defined to have their fronts at $u=0$ and
$v=0$, respectively, and thus to collide at  $u = v = 0$  (i.e. at $x^0 = 
x^1 = 0$). The two waves are {\it not} assumed to be impulsive, i.e. their 
energy density can have any (finite?) support at  positive $u$ and $v$, 
respectively. Space-time is thus naturally divided in four regions:

Region I, defined by $u,v<0$, is the space-time in front of the waves
before any
interaction
takes place. It is trivial Minkowski space-time:
\begin{equation}
ds_I^2=-2dudv+\sum \left(dx^i \right)^2 \; \; , B =0\; \; , \phi = \phi_0,
\end{equation}
with a constant perturbative dilaton
(${\rm exp}(\phi_0) \ll 1$). It will be convenient to fix the constant
in (\ref{phibar def}) as $-\phi_0$ so that, in region I, $\overline{\phi}= 
0$.

Region II, defined by $u>0$, $v<0$, is the wave coming from the left
before the interaction. Metric and dilaton depend only on
$u$ and therefore the field equations allow us to take
$F=0$ and write the ansatz:
\begin{equation}
ds_{II}^2=-2dudv+G_{ij}^{II}(u)dx^i dx^j \; \; , B_{ij}=B_{ij}^{II}(u)  
\; \; ,
\phi = \phi^{II}(u)\; .
\end{equation}
Similarly, region III, defined by $u<0$, $v>0$, represents
 the wave coming from the right
before the interaction. There,
\begin{equation}
ds_{III}^2=-2dudv+G_{ij}^{III}(v)dx^i dx^j \; \; ,
B_{ij}=B_{ij}^{III}(v) \; \; ,
\phi = \phi^{III}(v).
\end{equation}
Note that we have not assumed any special shape for $G$, so that
the results we give in this section will
hold whatever the (relative) polarization of the waves.

Finally, region IV ($u>0$, $v>0$) is the interaction region, with
\begin{equation}
ds_{IV}^2=-2e^F dudv+G_{ij}^{IV}dx^i dx^j,
\end{equation}
 and $F$, $G^{IV}$, $\phi^{IV}$ and
$B^{IV}$ are all functions of both $u$ and $v$.
Of course, the metric must be continuous along with its derivative on the 
boundary lines between the four regions. The same must be true for the 
dilaton
$\phi$ and the antisymmetric field $B$.

Let us begin by solving the equations in region II (and thus, by
trivial analogy, in region III).
The only non-trivial equation is the ``Virasoro constraint", eq.
(\ref{Vir}),
which (after momentarily dropping the subscripts II) reads
\begin{equation}
\ddot{\overline{\phi}}=\frac{1}{8}\mathrm{Tr}\left[\left( M^{-1}
\dot{M}\right)^2\right],\label{Region II1}
\end{equation}
where the dot indicates the derivative with respect to $u$.
The r.h.s. can be written as the sum of three terms:
\begin{equation}
\frac{1}{8}\mathrm{Tr}\left[\left( M^{-1}\dot{M}\right)^2\right]=
\frac{1}{4d}\left[\mathrm{Tr}\left(G^{-1}\dot{G}\right)\right]^2+\frac{1}{4}
\left\{\mathrm{Tr}\left[\left(G^{-1}\dot{G}\right)_{t}^2\right]-\mathrm{Tr}
\left[\left(
G^{-1}\dot{B}\right)^2\right]\right\},
\end{equation}
where $\left(G^{-1}\dot{G}\right)_{t}$ is the traceless part of
$\left(G^{-1}\dot{G}\right)$. The first trace on the r.h.s. can be
expressed in
terms of the dilaton and the shifted dilaton as
\begin{equation}
\frac{1}{4d}\left[\mathrm{Tr}\left(G^{-1}\dot{G}\right)\right]^2=\frac{1}{4d}
\left(\frac{d}{du}{\log\det G}\right)^2=
\frac{1}{d}\left(\dot{\phi}-\dot{\overline{\phi}} \right)^2.
\end{equation}
It is now useful to change  variable from $u$ to $\tilde{u}$, with
\begin{equation}
\frac{d}{du}=e^{-2\phi/d}\frac{d}{d\tilde{u}}.
\end{equation}
Then, eq. (\ref{Region II1}) becomes
\begin{equation}
e^{\overline{\phi}/d}\left(e^{-\overline{\phi}/d} \right)''=
-\frac{1}{d^2}\phi'^2-\frac{1}{4d}\left\{\mathrm{Tr}\left[\left(G^{-1}G'\right)
_{t}^2\right]-\mathrm{Tr}\left[\left(
G^{-1}B'\right)^2\right]\right\},\label{Region II2}
\end{equation}
where the prime denotes the derivative with respect to $\tilde{u}$.

It can be easly proved that all  terms on the r.h.s. of eq.
(\ref{Region II2}) are
negative definite, if we remember that $G$ is symmetric with positive
eigenvalues and $B$ is antisymmetric. Hence, for any non--trivial wave, 
$e^{-\overline{\phi}/d}$, which is constant and identically equal to 1 in
region I, must acquire a non-vanishing, negative, and never increasing
derivative in region II.
Thus, $e^{-\overline{\phi}/d}$  must vanish at some
finite $\tilde{u} = \tilde{u}^*$.
Returning now to the coordinate $u$, we see that, if the dilaton
 is bounded
(a necessary assumption
 if we want to use the tree-level effective action),
there exists a finite $u = u^*$ where
$e^{-\overline{\phi}/d}$ vanishes. Correspondingly, also $\det G$ vanishes,
and the metric of the transverse space will collapse to zero proper volume,
thereby producing a (coordinate) singularity.

It is not too difficult to estimate the order of magnitude of $u^*$ by  
multiplying eq. (\ref{Region II2}) by $e^{-\overline{\phi}/d}$ and by
 integrating it once after having gone back to the original  
$u$-variable. The result is an estimate of $u^{*-1}$ and is given in  
terms of an integral over $u$ of the
energy density per unit volume, i.e. in terms of the energy density per  
unit area.
One thus recovers the estimate $u^* \sim (G_N\rho)^{-1}$,
in which the appearance
of Newton's constant is somewhat fictitious since,
for waves of a given geometry,
the energy density scales like $G_N^{-1}$. The final result, reported in  
eq. (\ref{focald}), is a Lorentz-boost-invariant way of writing the  
same expression once both waves are considered simultaneously.

The same arguments can be repeated in region III, where $\det G$ has to
vanish at some finite  $v=v^*$ with $\overline{\phi} \rightarrow + \infty$.
These results  generalize
to any $D$ and to non--trivial antisymmetric fields, a well-known result
in $D=4$.

Let us finally analyse region IV, where the interaction between the
two waves occurs. We drop the subscript $IV$ from all functions.
We begin by using eq. (\ref{phibar}),
 which tells us that $e^{-\overline{\phi}}$ is the sum of a function
of $u$ and
a function of $v$.
The unique function of this type that matches the
boundary conditions with region I is
\begin{equation}
e^{-\overline{\phi}(u,v)}=
e^{-\overline{\phi}_{II}(u)}+
e^{-\overline{\phi}_{III}(v)}
-1.
\end{equation}
 We see that $e^{-\overline{\phi}(u,v)}$ must vanish on a
hypersurface joining  the coordinate singularities in
regions II and III and contained
 in
the region $u\leq u_0$, $v \leq v_0$ within region IV.

Let us now introduce two  new sets of coordinates
that simplify the analysis in region IV.
One set is of the light-cone type:
\begin{equation}
r = r(v) = 2 e^{-\overline{\phi}_{III}(v)  } - 1, \;\;\;
s = s(u) = 2 e^{-\overline{\phi}_{II}(u)  } - 1, \;\;
\end{equation}
while the second set is of the $t-x$ kind:
 \begin{eqnarray}
&& \xi= \frac {1}{2}  (r + s) = e^{-\overline{\phi}(u,v)}
\sim -t, \\
&& z = \frac {1}{2} (s - r) =
e^{-\overline{\phi}_{II}(u)}-e^{-\overline{\phi}_{III}(v)} .
\end{eqnarray}
Note that the coordinates $r,s$ run from $+1$ to $-1$ in  region IV with
their sum always positive except on the singular boundary where $ r+s = 0$.
Going from the original coordinates to either of the new sets
changes only the conformal factor of the 2-metric.
We may thus write, for instance,
\begin{equation}
ds_{IV}^2= - e^f d\xi^2+e^f dz^2 +G_{ij}dx^i dx^j,
\label{metricxi}
\end{equation}
where $f$ and $G$ are functions of $\xi$ and $z$. In $r$, $s$ coordinates,
 this becomes
\begin{equation}
ds_{IV}^2= - 2e^f dr ds +G_{ij}dx^i dx^j,
\end{equation}
where $f$ and $G$ are now functions of $r$ and $s$.
The
shifted dilaton is simply
\begin{equation}
\overline{\phi}=-\log \xi = - \log (r+s)/2  .
\end{equation}
Finally, in these coordinates, eq. (\ref{phibar}) becomes trivial and the 
only
equations to be solved, (\ref{Mevol},\ref{Vir}), become
\begin{eqnarray}
&& \partial_r \left( (r+s) M^{-1}
\partial_s M \right) + \partial_s \left( (r+s) M^{-1}
\partial_r M \right) = 0 \label{Mevol1},\\
&&   (r+s)^{-2} + (r+s)^{-1} \partial_r  F   +
\frac{1}{8}\mathrm{Tr}\left(\partial_r
M^{-1} \partial_r M \right) = 0, \\
&&  (r+s)^{-2} + (r+s)^{-1} \partial_s  F   +
\frac{1}{8}\mathrm{Tr}\left(\partial_s
M^{-1} \partial_s M \right) = 0.
\label{Vir1}
\end{eqnarray}

To end this section, let us discuss
 the relation between the above equations and those discussed in
 Refs. \cite{MV1,BMUV1}, where general solutions
 in the
homogeneous \cite{MV1} and quasi-homogeneous \cite{BMUV1}
case were derived. Here we are dealing with fields depending
also on
 one spatial coordinate. For this reason, we have built the matrix
$M$ by considering only the
remaining $d$ spatial dimensions, excluding the $g_{11}$ component. In the
approach to the singularity, we expect the time derivatives to dominate
over the spatial ones (this hypothesis can also be tested and
verified {\it a posteriori}, see section 6). Thus, close to the singularity, 
the
general solutions of \cite{MV1,BMUV1} should be recovered.
To see the explicit connection
between the equations of motion (\ref{Mevol1},\ref{Vir1}) and those
in \cite{MV1,BMUV1},
we have to complete the $M$ matrix with the missing rows and columns,
extending it to
a $2(d+1)$-dimensional matrix. We  must also add the $g_{11}$ component in 
the
determinant of the metric used to construct the shifted dilaton and go 
over to
cosmic time $dT=e^{f/2}d\xi$. After some simple algebra,
the equations of motion can be brought to a form identical to that of Refs. 
\cite{MV1,BMUV1},
 strongly suggesting that the asymptotic solutions of
\cite{BMUV1} will emerge near the singularity.

\section{Solutions with $B=0$}

The case of $B=0$ with parallel  polarized waves was discussed in Ref. 
\cite{FKV} for $d=2$. We shall obtain below
a generalization of their results to arbitrary $d$.
 For $B=0$,
$M$ reduces to
\begin{equation}
M=\left(
\begin{array}{cc}
G^{-1} & 0 \\
0 & G
\end{array}
\right).
\label{M}
\end{equation}
In each region, we choose a diagonal $G$ with $G_{ii}=e^{\lambda+\psi_i}$,
where $\sum \psi_i=0$ and $\lambda=\frac{1}{d}\log\det G$ goes to $-\infty$
at $u=u^*$ in region II and
at $v=v^*$ in region III.

The  equations of motion  follow easily from
eqs.  (\ref{Mevol1}) and(\ref{Vir1}). They read
\begin{equation}
\lambda_{,rs}+\frac{1}{2\left(r+s\right)}\left( \lambda_{,r}
+\lambda_{,s} \right)=0,
\end{equation}
\begin{equation}
\psi_{i,rs}+\frac{1}{2\left(r+s\right)}\left( \psi_{i,r}+\psi_{i,s}
\right)=0,
\end{equation}
\begin{equation}
f_{,r}+\frac{1}{\left(r+s\right)} - \frac{r+s}{4} \left( d \lambda_{,r}^2 +
\sum \psi_{i,r}^2 \right) =0, \label{fr eq}
\end{equation}
\begin{equation}
f_{,s}+\frac{1}{\left(r+s\right)} - \frac{r+s}{4} \left( d \lambda_{,s}^2 +
\sum \psi_{i,s}^2 \right) =0. \label{fs eq}
\end{equation}

We see that the equations for the $\psi_i$ and that for
$\lambda$ are
decoupled and can be solved separately. They are also formally the same, so
the solutions,  found by Szekeres \cite{Szekeres}, have the same structure:
\begin{eqnarray}
&& - (r+s)^{{\frac{1}{2}}} \psi_i\left(r,s\right) = \nonumber \\
&&  \int\limits_s^1
 ds' \left[(1+s')^{\frac{1}{2}} \psi_{i}(1,s') \right]_{,s'}
 P_{-\frac{1}{2}}\left[1+2\frac{
\left(1-r\right)\left(s'-s \right)}{\left(1+s' \right)
\left(r+s\right)}\right]
 \nonumber \\
&&  + \int\limits_r^1 dr' \left[(1+r')^{\frac{1}{2}} \psi_i(r',1)
\right]_{,r'}
P_{-\frac{1}{2}}\left[1+2\frac{
\left(1-s\right)\left(r'-r \right)}{\left(1+r' \right) \left(r+s\right)} 
\right].
 \label{psiformula}
\end{eqnarray}

The same expression holds for $\lambda$, with the obvious replacements.
In the above expressions, $P_{-\frac{1}{2}}(x)$ are  Legendre functions  
written
in standard notation. Once $\lambda$ is given, $\phi$ can be obtained from
eq. (\ref{phibar def}), since $\overline{\phi}$ is known.
Finally,
the function $f\left(r,s \right)$ is given by an integral along a curve 
joining the point $\left(r = s= 1 \right)$, where it vanishes, to the  
generic
point $\left(r,s \right)$:
\begin{equation}
f\left(r,s\right)=\int\limits_{\left(1,1\right)}^{\left(r,s\right)} \left[
f_{,r}dr + f_{,s}ds, \right]
\end{equation}
with $f_{,r}$ and $f_{,s}$ given by the r.h.s. of eqs. (\ref{fr eq})
and
(\ref{fs eq}) as functions of $\left(r,s \right)$ through the previously
determined $\phi$ and $\psi_i$.

We see that $\lambda$ and the $\psi_i$ are singular on the hypersurface
$\xi=r+s=0$. So, in a very general way, we  find that
 the collision of two plane
waves leads to a (curvature) singularity in the space-time whatever
 the number of
dimensions. Although we do not have, at the moment, such an explicit
solution for the general case, the discussion given in the previous section
makes us believe that a curvature singularity will always emerge
along the hypersurface $\xi=0$.

\section{Asymptotic approach to the singularity}

For the solutions (\ref{psiformula}), the asymptotic behaviour for
$\xi \rightarrow0$ is easily found by taking
the large-argument
limit of the Legendre function
\cite{Szekeres} (see also
Yurtsever \cite {Yurtsever}):
\begin{eqnarray}
&&\psi_i\left(\xi,z \right)\sim\epsilon_i\left(z\right)
\log \xi, \\
&&\lambda\left(\xi,z \right)\sim\kappa\left(z\right) \log \xi, \\
&&f\left(\xi,z \right)\sim a\left(z\right) \log \xi.
\end{eqnarray}
The coefficients multiplying the logarithm are functions of $z$,
whose range, on the singular surface, is $-1-+1$.
One easily finds
\begin{eqnarray}
&\epsilon_i\left(z\right)=&\frac{1}{\pi\sqrt{1+z}}\int\limits_z^1 ds
 \left[\left(1+s\right)^{\frac{1}{2}}\psi\left(1,s\right) \right]_{,s} 
\left(\frac{s+1}{s-z} \right)^{\frac{1}{2}}+ \nonumber \\
&&+ \frac{1}{\pi\sqrt{1-z}}\int\limits_{-z}^1 dr
 \left[\left(1+r\right)^{\frac{1}{2}}\psi\left(r,1\right) \right]_{,r} 
\left(\frac{r+1}{r+z} \right)^{\frac{1}{2}} \; ,\label{epsilon} \\
&a\left(z \right)=& \frac{1}{4}\sum\epsilon_i^2
\left(z\right)+\frac{d}{4}\kappa^2\left(z\right)-1 \; ,
\end{eqnarray}
with $\kappa\left(z\right)$ given by the same expression as $\epsilon_i$ 
with $\psi_i$ replaced by $\lambda$. The sum of the $\epsilon_i$ must be 
zero
according to the definition of the $\psi_i$.

The asymptotic form of the metric is
\begin{equation}
ds_{IV}^2=-\xi^{a\left(z\right)}d\xi^2+\xi^{a\left(z\right)}dz^2+
\xi^{\kappa(z)}\sum \xi^{
\epsilon_i\left(z\right)} \left(dx^i\right)^2 \; ,
\end{equation}
while
\begin{equation}
\phi \sim - \left(1+ \frac{d}{2}\kappa (z) \right) \log \xi \; .
\end{equation}

Going over to  cosmic time $\xi=t^{\frac{2}{a\left(z\right)+2}}$,
gives the metric in  Kasner form with exponents
\begin{eqnarray}
&&p_1\left(z\right)=\frac{a\left(z\right)}{a\left(z\right)+2} \label{p1},\\
&&p_i\left(z\right)=\frac{\kappa(z)+
\epsilon_i\left(z\right)}{a\left(z\right)+2} \label{pi}.
\end{eqnarray}
The following relations are immediately verified:
\begin{eqnarray}
&&\phi=\left(\sum\limits_{\alpha=1}^{D-1}
p_\alpha\left(z\right)-1\right)\log t,
\\
&&\sum\limits_{\alpha=1}^{D-1} p_\alpha^2\left(z\right)=1.
\end{eqnarray}

The behaviour of the fields near the singularity is thus of
Kasner type, modified, as usual, by the presence of the dilaton.
Note that, at the two tips of the singularity, $\epsilon_i$ and $\kappa$
diverge in such a way that the Kasner exponents near the tips
are simply $p_1 =1$, $p_i =0$. This corresponds to
 a (contracting) Milne-like metric
 which, being non-singular, nicely matches the non-singular
behaviour in regions II
and III. Away from these two points,
 $\kappa$ and $\epsilon_i$ can take any
value: it is easy to verify that the whole Kasner
sphere can be covered by appropriately choosing  the initial data.

For generic $z$, the collision
 leads to a singularity showing all the
characteristics of a cosmological singularity. The Kasner exponents depend 
on
 $z$ and therefore the behaviour of the metric will depend on the
point of the singular hypersurface that we approach as $\xi \rightarrow 
0$.
It is therefore interesting to study the
signs of the Kasner exponents for different choices of the coefficients
$\epsilon_i$ and $\kappa$. In particular, we would like to see
whether, far from the tips
of the singularity, inflation may take place.

By the definition of $a(z)$, we see that the denominators of the Kasner
exponents (\ref{p1}) and (\ref{pi}) are always positive. Concerning
the numerators, we note that
$a\left(z\right)$ is a quadratic form in
$\epsilon_i$ and $\kappa$. The equation $a\left(z\right)=0$ thus defines
an ellipsoid in the space $\left(\epsilon_i;\kappa\right)$ with  $p_1$  
positive outside of this ellipsoid and negative inside.
For each $i$ the equation $p_i=0$ defines the plane
$\kappa+\epsilon_i=0$ passing through the centre of the former
ellipsoid. The $p_i$'s are positive on the semispace containing the
positive
$\kappa$ semiaxis. Finally, we have to remember the constraint on the sum 
of
the $\psi_i$, which becomes $\sum \epsilon_i=0$ in the asymptotic
regime.

In conclusion, we see that  there indeed exists
 a region where all the Kasner exponents are
negative. In fact, if we stay inside the ellipsoid on the side where
$\kappa<0$, and take all $\epsilon_i$  close enough to zero, all the
exponents are  negative and we have inflation in all directions.
 In the opposite situation (i.e. outside the
ellipsoid, with $\kappa>0$ and $\epsilon_i$ close to zero) all exponents 
are
positive and we have contraction in all directions.
All intermediate cases are also possible as we can let any number of
$\epsilon_i$ be large and negative and  take
 at least one $\epsilon_i$
large enough and positive to balance the others.

\section{Colliding waves with antisymmetric field}

Although we have written the equations in a manifestly $O(d,d)$-covariant
form, we have only managed, so far, to find solutions in region IV for
 vanishing
antisymmetric field. We can perform, however, $O(d,d)$ boosts on our
solutions
to introduce it.
This procedure can be worked out in any number of the dimensions.
Our purpose in this section is to discuss the main features of the
cosmologies
obtained by this procedure and to clarify the role of the antisymmetric 
field,
particularly in connection with the work of Ref. \cite{DH1}.

Let us write the transformation matrix $\Omega$ in terms of four
$d$-dimensional blocks
\begin{equation}
\Omega=\left(
\begin{array}{cc}
P & R \\
Q & S
\end{array}
\right).
\end{equation}
The constraint (\ref{O(d,d) definition}) reads
\begin{eqnarray}
&& P^T Q+Q^T P = 0, \label{PQ relation}\\
&& R^T Q+ S^T P =I_d, \\
&& P^T S+ Q^T R = I_d, \\
&& R^T S + S^T R =0.
\end{eqnarray}
From eq. (\ref{Transformation for M}), with $M$ in the form (\ref{M}), we 
obtain the inverse of the new metric $G'$
\begin{equation}
G'^{-1}=P^T G^{-1} P+ Q^{T} G Q
\end{equation}
and extract also the antisymmetric field
\begin{equation}
B'=\left( R^T G^{-1} P+S^T G S \right) G'.
\end{equation}

It is interesting to investigate the asymptotic behaviour of the new
cosmologies obtained this way.
We try to be very general, considering an initial metric that is
diagonal and has the Kasner form
\begin{equation}
G=\mathrm{diag}\left(t^{\lambda_1}, \ldots, t^{\lambda_d}\right),
\label{G}
\end{equation}
with the $\lambda_i$ ordered so that $i<j$ implies that $|\lambda_i| \geq
|\lambda_j|$.
In section 5 we showed that the asymptotic approach to the singularity of the
colliding waves metric is of this type, if we let the
$\lambda_i$
depend on $z$.

When we perform the $O\left( d,d\right)$ transformation, the new metric 
$G'$ is no longer diagonal. However, we can diagonalize it by a
step-by-step procedure.
Exploiting the fact that $G$ has the form (\ref{G}), the eigenvalue
equation
for $G'^{-1}$ takes the following asaymptotic form
\begin{equation}
\sum\limits_{b} \sum\limits_i \left(P_{ia} t^{-\lambda_i} P_{ib}+ Q_{ia}
t^{\lambda_i} Q_{ib} \right) v_b = \alpha v_a.
\end{equation}

Assuming, in a first instance, that all $\lambda_i$ are non-vanishing,
 we define
two new matrices $U$ and $U'$, whose components are
\begin{eqnarray}
&& U_{ij}=\left\{ \begin{array}{l}
P_{ij} \; \; \; $if $ \lambda_i>0  \\
Q_{ij} \; \; \; $if $ \lambda_i < 0
\end{array} \right. ,\\
&& U'_{ij}=\left\{ \begin{array}{l}
Q_{ij} \; \; \; $if $ \lambda_i>0 \\
P_{ij} \; \; \; $if $ \lambda_i < 0
\end{array}\right. .
\end{eqnarray}
The dominant term in the eigenvalue equation as
$t\rightarrow 0$ becomes
\begin{equation}
\sum\limits_{b} U_{1a} t^{-|\lambda_1|} U_{1b} v_b = \alpha v_a.
\end{equation}
This is the eigenvalue equation for the projector to $U_1$, i.e.
the first eigenvalue of $G'^{-1}$ is $\alpha= |U_{1}|^2
t^{-|\lambda_1|}$
with eigenvector $v_a=U_{1a}$. Since the vectors
orthogonal to $U_1$ annihilate the dominant term we just considered, in 
order to   find the other eigenvalues and the corresponding eigenvectors, 
we  consider the successive sub-dominant terms in the original eigenvalue 
equation  in the space orthogonal to $U_1$.

The next term is
\begin{equation}
\sum\limits_{b} U_{2a} t^{-|\lambda_2|}
U_{2b} v_b = \alpha v_a.
\end{equation}
The eigenvector of this term is
$v=U_2$: however, in order  to annihilate the  dominant term, we have to 
take the component of $U_2$ (asymptotically)  orthogonal to $U_1$ as the 
second eigenvector. The corresponding eigenvalue is $\alpha= \left(
|U_2|^2- \frac{\left( U_1 \cdot U_2 \right)^2 }{|U_1|^2} \right)
t^{-|\lambda_2|}$. The next eigenvector must now be taken in the space 
orthogonal to $U_1$ and $U_2$, and so on. In this way, we can build the 
complete system of eigenvectors of $G'^{-1}$, combining the rows of $U$, 
which are rows of the two matrices $P$ and $Q$. The $d$ eigenvalues are 
then proportional to $t^{-|\lambda_i|}$  and therefore they are all
diverging as $t \rightarrow 0$. It is easy to extend this discussion to 
the case of some vanishing $\lambda_i$. This clearly implies
asymptotically vanishing eigenvalues for $G$.

In order to complete our discussion, we have to mention some possible  
exceptions to the above-mentioned situation. Not always, the  rows of $U$, 
used to build the eigenvectors, form an independent set of  vectors. For a 
particular class of $O(d,d)$ transformations, having $\det U=0$, the
result is different. It is easy to understand the modifications: each time 
we face a $U_i$ linearly dependent on the previous rows, we just ignore it 
and proceed with the algorithm to the next term in $G'^{-1}$. When we have 
exhausted the first $d$ dominant terms, we will still have to find one 
eigenvalue (having skipped one term  in the diagonalization). Taking the 
next term of $G'^{-1}$,  in the form
\begin{equation}
U'_{da}
t^{|\lambda_d|} U'_{db},
\end{equation}
we can find the missing eigenvalue.
In conclusion, we can see that, if the rank of $U$ is $s \leq d$, then  
$G'^{-1}$ will have $s$ diverging  and $d-s$ vanishing eigenvalues. Note 
that, even if the  matrix $P$ is chosen freely, the matrix $Q$ is related 
to $P$ by eq. (\ref{PQ relation}). We  have to check whether, because
of this relation, the determinant of $U$ is  necessarily vanishing for 
some initial values of the Kasner exponents.

First, we observe that (\ref{PQ relation}) implies
\begin{equation}
\det P
\det Q= (-1)^d \det Q \det P.
\end{equation}
Therefore, for odd $d$,
either the determinant of $P$  or the determinant of $Q$  vanish. We may 
now distinguish two classes of $O(d,d)$ transformations. The identity
transformation belongs to those giving a vanishing $\det Q$, and the same 
is true for  all transformations that  can be reduced to the identity in a 
continuous way. For metrics that are  inflating in all directions  before 
the transformation, $U=Q$ and the transformations having $\det Q=0$,
instead of  turning  all eigenvalues to contracting ones, leave at least 
an inflating. Conversely, for metrics contracting in all dimensions,
$U=P$ and transformations with $\det P=0$ cannot leave all the eigenvalues 
unchanged, but induce inflation in  at least one direction.

Consider, as the last peculiar case, the one in which  the  exponents in 
the initial metric are all positive except one. Then it is possible to 
show that the constraint (\ref{PQ relation}) forces  $\det U =0$ if $P$ is 
non-singular.  The same argument can be repeated for metrics inflating in 
all directions except  one. In this case, $U$ is composed by $d-1$ rows 
from $Q$ and one from $P$. If $Q$ is non-singular, then $\det U=0$ and  
again we have one inflating dimension left.

Summing up, we can say that cosmologies with antisymmetric field that can 
be obtained, by an $O(d,d)$ transformation, from a metric having Kasner 
behaviour near the singularity, generally contract in all dimensions. In 
even dimensions, at least one  inflates when the original metric has just 
one inflating or just one contracting dimension. In odd dimensions, if the 
$O(d,d)$ transformation has $\det Q=0$, then, starting   from a full
inflating metric or from a metric with one inflating dimension, we are 
left with one inflating dimension. If $\det P=0$, then one dimension
inflates when we start from a full contracting metric or from  a metric 
having just one contracting dimension.

We should also recall that none of the $O(d,d)$ transformations  
affects the $g_{11}$ component of the metric, which, therefore, can  
either inflate or contract.

The results  of Ref. \cite{CFLT} are compatible with ours, since,  for 
$d=2$, our statements,  when the determinant of $G$ inflates,  can be
summarized as a change of sign in $\lambda$ while $\psi$ is left invariant.

\section{Conclusions}

In this paper, extending previous work \cite{FKV},  we have modelled the 
onset of pre-big bang inflation, from asymptotically trivial initial
conditions \cite{BDV},  as the result of the collision of two plane waves,  
made of gravitons, dilatons, and Kalb--Ramond massless particles,  in any 
number $D=d+2$ of space-time dimensions. We showed that the evolution of 
the system is described  in terms of a compact and elegant set of
$O(d,d)$-invariant equations and that properties of the solutions can be 
studied in full generality in three of the four regions defined by the 
planar-collision problem.  This already enables us to argue that the  
formation
of a curvature singularity in the future (to be identified with the big 
bang) is generic.

However, so far, we have not been able  to solve the general problem
analytically in the fourth, and most interesting, region, except for the 
case of dilatonic and parallel-polarized gravitational waves.
Nonetheless, using  $O(d,d)$ transformations, we were able to construct 
new solutions containing the KR field and to discuss the physically
relevant properties of these new solutions. The main conclusions that
appear to emerge are the following:
\begin{itemize}
\item While the
formation of a singularity is generic,  the existence of inflating regions 
near the singular surface is only generic  in the absence of the KR form. 
\item The KR form, at least when generated from $O(d,d)$ transformations,  
tends to generate contraction rather than expansion, in  agreement with 
other results \cite{PBB} and arguments \cite{DH1}.
\item Since our
equations appear to reduce,  near the singularity, to those studied
previously \cite{MV1,BMUV1} in the homogeneous (or quasi-homogeneous)
case, it looks very likely that the above behaviour will persist in the 
general solution.
\end{itemize}

Eventually, one may be led to the conclusion that the most generic  APT 
initial conditions (i.e. those containing  all possible kinds of waves in 
the initial state) can hardly produce such a flat, homogeneous and
isotropic Universe to  dispense us  completely from the more standard kind 
of potential-energy-driven  post-big bang inflation. Actually, it is very 
likely that, after exit from pre-big bang inflation, the dilaton and other 
moduli will find themselves  dispaced from the minima of their
non-perturbatively-generated potentials and that further inflation will 
result from their rolling down towards them. A similar conclusion  
seems to follow from  completely different, more phenomenological  
arguments,  i.e.
from the recent data analysis  of CMB anisotropies at small angular scales 
\cite{Boomerang}, which appears to confirm the need for adiabatic
perturbations of the kind naturally provided by potential-driven
inflation,  but absent in the PBB scenario.

If so, we will have to back up from the early claims that the PBB scenario 
can replace altogether  standard inflation, and settle instead for  a
complementary role it would play in providing the  initial conditions that 
standard inflation badly needs, and in ``explaining", from more natural 
and generic initial conditions, the most mysterious event  in the entire 
life of our Universe, the big bang.

\begin{centerline}
{\bf Note Added}
\end{centerline}

While this paper was being written, we received a new paper by Damour and 
Henneaux \cite{DH2} containing a discussion of the outcome of $O(d,d)$ 
transformations on Kasner-like solutions. Their results, obtained  with a 
different diagonalization procedure,
agree with those discussed in our section 6.

\bigskip

\begin{centerline}
{\bf Acknowledgements}
\end{centerline}
We are grateful to Maurizio Gasperini for help during the early
 stage of this work and to  Thibault Damour for discussions
concerning his recent work, Refs. \cite{DH1,DH2}.
V.B. was supported by fund ex 60\% D.P.R. 382/80, MURST,
Italy. He also wishes to thank
the Theory Division at CERN for hospitality while this work was carried 
out.


\begin{thebibliography}{}

\bibitem{PBB} J.E. Lidsey, D. Wands and E.J. Copeland,
hep-th/9909061, to appear in Phys. Rep.;
 G. Veneziano, hep-th/0002094, and references therein.

\bibitem{TW} M. Turner and E. Weinberg, Phys. Rev. { D56} (1997) 4604;\\ 
 N. Kaloper, A. Linde and R. Bousso, Phys. Rev. { D59} (1999) 043508; \\
A. Buonanno, K.A. Meissner, C. Ungarelli
and G. Veneziano, Phys. Rev. { D57} (1998) 2543;
M. Gasperini, hep-th/0004149.

\bibitem{BDV} A. Buonanno, T. Damour and G. Veneziano, Nucl. Phys. B543 
(1999) 275.

\bibitem{Chr} D. Christodoulou, Comm. Math. Phys. 105 (1986) 337,
	Comm. Pure Appl. Math. 54 (1991) 339 and
	Comm. Pure Appl. Math. 56 (1993) 1131.

\bibitem{FKV} A. Feinstein, K.E. Kunze, M.A. V{\'a}zquez--Mozo,
hep-th/0002070.

\bibitem{DH1} T. Damour and M. Henneaux, hep-th/0003139.

\bibitem{BKL}  V.A. Belinskii, I.M. Khalatnikov and E.M. Lifshitz, Adv. 
Phys.
19 (1970) 525.

\bibitem{Szekeres} P. Szekeres, J. Math. Phys. 13 (1972) 286.

\bibitem{MV1} K.A. Meissner and G. Veneziano, Phys. Lett. B267 (1991) 33 and
 Mod. Phys. Lett. A6 (1991) 37.

\bibitem{Sen} A. Sen, Phys. Lett. B271 (1991) 295; \\
S.F. Hassan and A. Sen, Nucl. Phys. B375 (1992) 103.

\bibitem{GV} M. Gasperini and G. Veneziano, Phys. Lett. B277
(1992) 256.

\bibitem{Tseytlin}
G. Veneziano, Phys. Lett. B265 (1991) 287;\\
A.A. Tseytlin, Mod. Phys. Lett. A6 (1991) 1721; \\
	A.A. Tseytlin and C. Vafa, Nucl. Phys. B372 (1992) 443.

\bibitem{Mah Sch} J. Maharana, J.H. Schwarz, Nucl. Phys. B390 (1992) 3.

\bibitem{BMUV1} A. Buonanno et al., Ref. \cite{TW}.

\bibitem{Yurtsever} U. Yurtsever, Phys. Rev. D37 (1988) 2803.

\bibitem{CFLT} D. Clancy, A. Feinstein, J.E. Lidsey, R. Tavakol, Phys. Rev.
D60 (1999) 043503.

\bibitem{Boomerang} A.E. Lange et al., astro-ph/0005004; \\
   A. Balbi et al., astro-ph/0005124.

\bibitem{DH2} T. Damour and M. Henneaux, hep-th/0006171.

\end{thebibliography}
\end{document}